\documentstyle[12pt,twoside,fleqn,espcrc1]{article}

\def\Journal#1#2#3#4{{#1} {\bf #2}, #3 (#4)}

\def\NPB{{ Nucl. Phys.} B}
\def\NPA{{ Nucl. Phys.} A}
\def\PLB{{ Phys. Lett.}  B}
\def\PRL{ Phys. Rev. Lett.}
\def\PRD{{ Phys. Rev.} D}
\def\PRC{{ Phys. Rev.} C}
\def\ZPC{{ Z. Phys.} C}

\def\be{\begin{equation}}
\def\ee{\end{equation}}
\def\bea{\begin{eqnarray}}
\def\eea{\end{eqnarray}}

\newcommand{\AmS}{{\protect\the\textfont2
  A\kern-.1667em\lower.5ex\hbox{M}\kern-.125emS}}

\title{QCD Properties of Hot/Dense Matter}

\author{Su Houng Lee\address{Department of Physics, Yonsei University, 
\\ Seoul 120-749, Korea}%
        \thanks{
This work was supported in part by the Korean Ministry of Education through 
grant no. BSRI-97-2425 and  by the KOSEF through grant no. 971-0204-017-2 and the CTP at Seoul National University.}
       }

\input{epsf}

\begin{document}
\maketitle

\begin{abstract}
In the introduction, we will discuss the symmetries of QCD, its breaking in 
the physical vacuum and its restoration at finite temperature and/or
 density.  Then we will discuss what exact relations can be made between 
chiral symmetry restoration and the imaginary part of the vector vector 
correlation function, which is directly related to the dilepton spectrum
 coming from a hot and/or dense medium.   
Finally, we will focus on the vector meson pole appearing in the imaginary
part and discuss results on how the shape changes using  QCD sum rules in 
medium.  
 In particular, we will discuss three 
characteristic changes of the vector meson peaks; namely, shift of the 
peak position, increase of its width and the dispersion effect due to 
the breaking of Lorentz invariance in medium.
\end{abstract}

\section{INTRODUCTION}

 At zero temperature and/or baryon density, the chiral symmetry is 
spontaneously broken 
 $\langle \bar{q} q \rangle \neq 0 $ and  the hadronic world has only  
 $SU(N)_V \times U(1)_V$ symmetry.
However, if the 
temperature or baryon density increases, there will be a critical boundary
above which the broken chiral symmetry gets restored $ \langle \bar{q} q 
\rangle =0$ and the hadronic matter changes to a quark gluon plasma (QGP) 
state.  This QGP state will have $SU(N)_L \times SU(N)_R \times U(1)_V$ 
symmetry, which is also the symmetry of QCD at the quantum level.

\subsection{Symmetry of QCD}\label{subsec:qcd}

The Lagrangian of QCD is;
\be
{\cal L}_{QCD}= -\frac{1}{4} F^a_{\mu \nu}F^a_{\mu \nu} - \sum_f \bar{q}_f
 (i \slash \hspace*{-0.25cm} D + M_f) q_f
\ee
where $F^a_{\mu \nu}=\partial_\mu A^a_\nu-\partial_\nu A^a_\mu +ig f^{abc}
A^b_\mu A^c_\nu$ and $A^a_\mu$ represents the spin 1 gauge field with color 
index $a$.  $q_f$ represents the spin $\frac{1}{2}$ matter field which has 
left and right chirality components,  $ q_{L,R}=\frac{1}{2} (1 \mp \gamma_5)
q$, and the index $f$ represents the $N$ number of flavors such as the 
 $u,d,s \cdot \cdot $.  
In the chiral limit $(M \rightarrow 0)$,  ${\cal L}_{QCD}$ is symmetric under
 $U(N)_L \times U(N)_R$, i.e., the lagrangian is symmetric under the 
following left right transformation.

\be
\left( \begin{array}{c} u \\ d\\ s \\ \cdot \end{array} \right)
\rightarrow e^{iQ} 
\left( \begin{array}{c} u \\ d\\ s \\ \cdot \end{array} \right)
\ee
where $Q=\frac{1-\gamma_5}{2}(L_0 + L_i \lambda^i) +
\frac{1+\gamma_5}{2}(R_0 + R_i \lambda^i)$.  Under this transformation, the 
chirality does not mix ($q_{L,R} \rightarrow q_{L,R}$).   

However, at the quantum level, due to the axial anomaly, the $U_A(1)$ 
symmetry ($Q=\gamma_5(R_0-L_0)$) is broken and QCD is symmetric only under
\be
SU(N)_L \times SU(N)_R \times U(1)_V.
\ee
This is also the symmetry of the QGP state.

\subsection{Spontaneously broken chiral symmetry}

In the hadronic world, the QCD symmetry at quantum level is further
broken down spontaneously to the following group.
\be
SU(N)_V \times U(1)_V
\ee
As a consequence the QCD vacuum and physical excitations are not symmetric 
under $Q=\gamma_5 Q^a \lambda^a$ transformation.  This is manifest from
looking at various order parameters of chiral symmetry.

 First, consider the vacuum state.  The chiral partner of any quark composite
operator $ Op$ is obtained by taking its commutator with the chiral charge 
operator $F^a_5=\int d^3x \bar{q} \gamma_0 \gamma^5 \frac{\lambda^a}{2}q $.  Hence the 
chiral partner of $ \bar{q} q $ is $ \bar{q} \gamma_5 
\frac{\lambda^a}{2}q $.    Since the vacuum expectation value of the latter 
vanishes identically,  $\langle \bar{q} q \rangle \neq 0$ shows that chiral
symmetry is broken in the vacuum.   The fact that chiral symmetry is 
spontaneously broken is also manifest in the presence of massless Nambu-Goldstone Bosons ( the $\pi$'s), which couple to the broken axial 
current $\langle 0| A_\mu^5 | \pi \rangle \sim f_\pi$.

As can be seen from table 1, 
the broken chiral symmetry is also manifest in the mesonic  
excitation, such as $m_\rho \neq m_{a_1}$, and baryonic excitations, such 
as $m_P \neq  m_{S_{11}}$\cite{Oka,Hung}.

\begin{table}[hbt]
\newlength{\digitwidth} \settowidth{\digitwidth}{\rm 0}
\catcode`?=\active \def?{\kern\digitwidth}
\caption{Chiral  order parameters and evidence for its breaking in hadronic world. 
Here $V_\mu^a= \bar{q} \gamma_\mu \frac{\lambda^a}{2} q, A_\mu^a 
= \bar{q} \gamma_\mu \gamma_5 \frac{\lambda^a}{2} q$. $\Psi_-$ is the chiral 
partner of $\Psi_+$.  
 \label{tab:order}}
\begin{tabular*}{\textwidth}{@{}l@{\extracolsep{\fill}}rr}
\hline
 \raisebox{0pt}[13pt][7pt]{Order parameter} &
 \raisebox{0pt}[13pt][7pt]{Hadronic world} 
\\
\hline
 \raisebox{0pt}[13pt][7pt]{ $ \langle \bar{q} q \rangle$ } &
 \raisebox{0pt}[13pt][7pt]{ $ \langle \bar{q} q \rangle \neq 0, 
f_\pi^2m_\pi^2=-\langle m \bar{q} q \rangle $ }
\\
 \raisebox{0pt}[13pt][7pt]{ $ \langle V_\mu^a(x) V_\mu^a(0) \rangle
 -\langle A_\mu^a(x) A_\mu^a(0) \rangle $ } &
 \raisebox{0pt}[13pt][7pt]{ $m_\rho(770) \neq m_{a_1} (1250)$} 
\\ 
 \raisebox{0pt}[13pt][7pt]{ $ \langle \Psi_+(x) \Psi_+(0) \rangle
 -\langle \Psi_-(x) \Psi_-(0) \rangle$ } &
 \raisebox{0pt}[13pt][7pt]{ $m_P(938) \neq m_{S_{11}}(1535)$}\\ 
\hline
\end{tabular*}
\end{table}

\subsection{Chiral symmetry restoration at high temperature and/or density}

As seen from lattice gauge theory calculations, the broken chiral symmetry
will be restored at high temperature.   The  transition will become smooth 
as we increase the current quark mass.   The actual order of the phase 
transition at the physical value of the quark masses for the $u,d,s$ quarks 
are still controversial.  The result using the Staggered 
fermions\cite{Columbia} shows a smooth transition at the physical value of 
the quark masses.  However, the result using the Wilson fermions\cite{Iwasaki}
 shows a first order transition.

\begin{figure}[htb]
\begin{minipage}[t]{80mm}
\epsfxsize=300pt
\leftline{\epsfbox{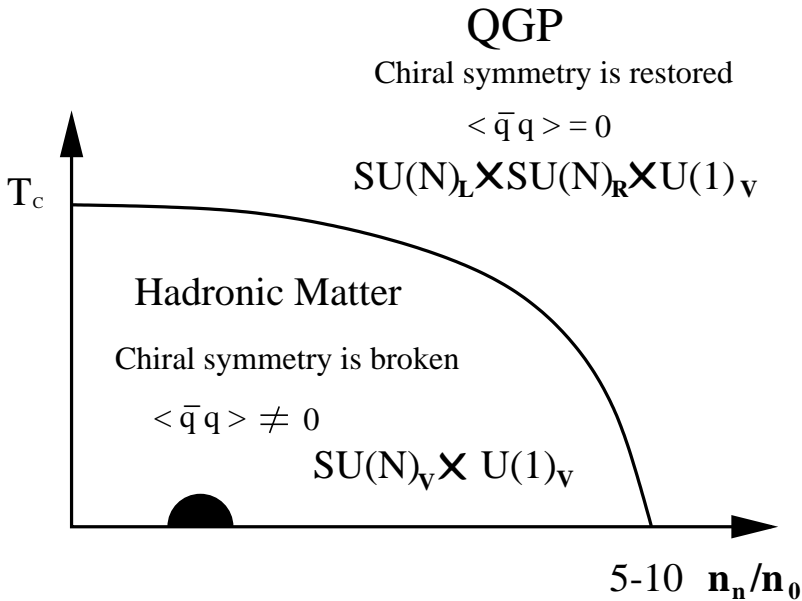}}
\caption{Phase diagram of QCD.}
\label{fig:large}
\end{minipage}
\end{figure}

At present, the 
lattice calculation at finite density suffers a technical difficulty 
due to complex fermion determinant.  Nevertheless,  many model calculations, 
such as instanton models\cite{Schafer}, NJL model, Random Matrix model\cite{random}
suggest that chiral symmetry will also be restored at finite baryon density.  
In fact, already at nuclear matter density, there exist model independent
relations that shows that the chiral order parameter will be reduced to about
 80 \% of its vacuum value.  Therefore, any physical effect related to chiral 
symmetry restoration will also be partially visible in nuclear matter.  

\subsection{ $U_A(1)$ restoration at high temperature and/or density}

 Present lattice calculations shows that the $U_A(1)$  symmetry is not restored
above the chiral phase transition\cite{Bernard97,Boyd}.  This is reasonable because the $U_A(1)$ breaking effect comes from a totally different mechanism from that of chiral symmetry breaking.
Nevertheless, 
the present lattice calculations were performed with $N=2$ flavors 
and studied the $U_A(1)$ order parameter constructed out of two quark 
bilinears (2-point function).  It would be interesting to perform 
similar calculations with larger number of flavors.  The reason, which is
valid above $T_c$, is the 
following.  
The $U_A(1)$ breaking effect comes from topologically non-trivial gauge field
configurations.    Its contribution to the partition function 
has measure zero in the chiral limit, i.e. its contribution is 
proportional to the  $M^N \rho_{inst}$, where $N$ is the number of quark
flavors having current quark mass $M$, and $\rho_{inst}$ is the pseudo-instanton
density\cite{tHooft}.   For $U_A(1)$ order parameter, constructed out of $n$ quark 
bilinears, the breaking contribution is proportional to 
 $M^{N-n} \rho_{inst}$.   
This implies, that in the chiral limit, after chiral symmetry
restoration,  the $U_A(1)$ breaking effect shows up only when $n \geq N$\cite{LH96}.

\begin{table}[hbt]
\catcode`?=\active \def?{\kern\digitwidth}
\caption{ $U_A(1)$  breaking operators and its breaking contribution from
the partition function.
Here $\pi^a= \bar{q} \gamma_5 \frac{\lambda^a}{2} q, \delta^a 
= \bar{q} \frac{\lambda^a}{2} q$.  $N$ is the number of quark flavors with 
mass $M$.  $\rho_{inst}$ is the pseudo-instanton density.
 \label{tab:u1}}
\begin{tabular*}{\textwidth}{@{}l@{\extracolsep{\fill}}rr}
\hline
 \raisebox{0pt}[13pt][7pt]{Breaking operator} & 
 \raisebox{0pt}[13pt][7pt]{ Breaking contribution}\\
\hline
 \raisebox{0pt}[13pt][7pt]{$\langle  \bar{q} q \rangle$  } & 
 \raisebox{0pt}[13pt][7pt]{ $M^{N-1} \rho_{inst}$}
\\
 \raisebox{0pt}[13pt][7pt]{$\langle \pi^a(x) \pi^a(0) \rangle 
  - \langle \delta^a(x) \delta^a(0) \rangle$ } & 
 \raisebox{0pt}[13pt][7pt]{ $M^{N-2} \rho_{inst}$ }
\\ 
 \raisebox{0pt}[13pt][7pt]{ n-point function} & 
 \raisebox{0pt}[13pt][7pt]{ $M^{N-n} \rho_{inst}$ }
\\ 
\hline
\end{tabular*}
\end{table}

\section{ OBSERVING CHIRAL SYMMETRY RESTORATION}

Among the order parameters of chiral symmetry restoration, such as those
discussed in Table. 1, let us concentrate on  the difference between the 
vector and axial vector two point functions.  Using the dispersion relation,
it can also be related to the 
difference between the imaginary parts.  
\begin{eqnarray}
\label{vadiff}
\Pi^V_{\mu \mu}(q) -\Pi^A_{\mu \mu}(q)
= \frac{1}{\pi} \int ds \left( \frac{ {\rm Im} \Pi^V (s)}{s-q^2} - 
\frac{ {\rm Im} \Pi^A (s)}{s-q^2} \right),
\end{eqnarray}
where,
\begin{eqnarray}
\label{pol}
\Pi^V_{\mu \nu}  =  \int d^4x e^{iqx}  \langle V_\mu(x) V_\nu (0) \rangle 
~,~~~
\Pi^A_{\mu \nu}   =  \int d^4x e^{iqx}  \langle A_\mu(x) A_\nu (0) \rangle .
\end{eqnarray}
 The imaginary part of the vector correlator can be observed from 
the hadronic part of the 
cross-section  $e^+ +e^- \rightarrow  ~ {\rm even}~ \pi$ Fig.2 (a).  
This  has a sharp rho meson peak (770 MeV) followed by resonances, 
which add up like a continuum.  
  The imaginary part of the axial correlator can be observed from 
 the hadronic part of the decay 
 $\tau \rightarrow ~ {\rm odd} ~\pi$ Fig.2 (b).   This 
part has the broad $a_1$ peak (1260 MeV).

\begin{figure}[htb]
\begin{minipage}[t]{80mm}
\epsfxsize=400pt
\leftline{\epsfbox{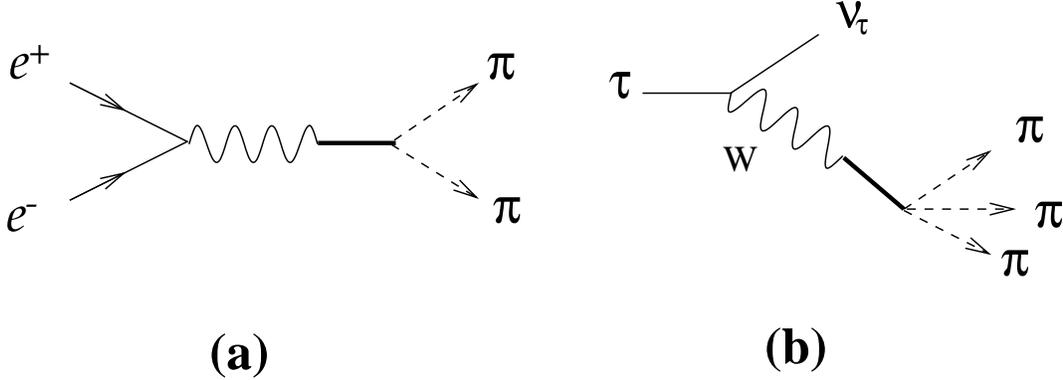}}
\caption{a)$e^+ +e^- \rightarrow  ~ {\rm even}~ \pi$, b)$\tau 
\rightarrow ~ {\rm odd} ~\pi$.}
\label{fig:largenenough}
\end{minipage}
\end{figure}
The two
imaginary parts are clearly different and is one of the  experimental signature that
 chiral symmetry is broken in the vacuum.

In the vacuum, due to current conservation and Lorentz invariance, the
polarization functions in Eq. (\ref{pol}) has only one independent
function, which is a function of the four momentum $q^2=\omega^2-{\bf q^2}$.
At finite temperature or density, the Lorentz invariance is 
broken and the polarization functions will have a transverse ($\Pi_T$)
and a longitudinal ($\Pi_L$)
component (with respect to ${\bf q}$), which will depend on
 both the energy $\omega$ and three momentum ${\bf q}$.

\begin{eqnarray}
\Pi_{\mu\nu}(\omega,{\bf q})=\Pi_T q^2 {\rm P}^T_{\mu\nu}+ \Pi_L q^2 
 {\rm P}^L_{\mu\nu},
\label{pol2}
\end{eqnarray}   
where for $q=(\omega,{\bf q})$ and medium at rest, we have, 
$ 
{\rm P}^T_{00}  =  {\rm P}^T_{0i}={\rm P}^T_{i0}=0,~~
{\rm P}^T_{ij}  =  \delta_{ij}-{\bf q}_i {\bf q}_j/{\bf q}^2,$ and $
{\rm P}^L_{\mu\nu}  =  (q_\mu q_\nu/q^2-g_{\mu\nu}- {\rm P}^T_{\mu\nu}) $.  
 $\Pi_L= \Pi_T$ only when ${\bf q}=0$.

Above the critical temperature or density, chiral symmetry is restored and
 ${\rm Im} \Pi^V_{L,T}={\rm Im}\Pi^A_{L,T}$, 
 which at sufficiently high temperature or density 
can be calculated using perturbative QCD\cite{BS86}. 
At high temperature or density, the poles disappear and become 
part of the continuum. 
Now the question is how would the poles behave and approach the continuum from  
below the critical boundary?

In fact, one can  observe ${\rm Im}\Pi^V$ from a hot or dense matter by 
looking at the inverse process $e^+ + e^- \rightarrow {\rm even} ~\pi$.  
That is, by looking at the dileptons coming out from the medium\cite{Pi82}.  Since, the dileptons interact weakly
in the strong interaction time scale, once they are produced, they will 
carry the information about  ${\rm Im}\Pi^V$  directly to detectors.  
This would be true in the time scales involved in Relativistic Heavy Ion collisions or experiments of vector meson production  inside a Heavy Nucleus
planned at GSI and KEK.  
Unfortunately, the inverse process of $\tau \rightarrow {\rm odd}~\pi$, that is $\tau +\nu$ coming 
out from the medium  is not experimentally observable and one can not 
directly observe ${\rm Im}\Pi^A$.    So the best we can do to observe chiral
symmetry restoration (at least in heavy ion collisions) is to 
experimentally observe ${\rm Im} \Pi^V$  (especially the characteristic 
changes in the peak) and do QCD model calculations to 
find its relation to chiral symmetry restoration.

The dilepton production rate from a hadronic system at finite temperature
is given  as follows.

\begin{eqnarray}
\label{prod1}
\frac{dR }{dM^2}= \frac{2 \alpha_{em}^2}{3 \pi^2} \int d k \frac{k^2}{
 \omega } \frac{1}{M^2}  [ {\rm Im}\Pi_L(\omega,k) + 
2 {\rm Im}\Pi_T(\omega,k) ] n(\omega)
\end{eqnarray}
where, $\omega^2=k^2+M^2$ and $n(\omega)$ is the thermal Boltzmann factor. 
Here, ${\rm Im}\Pi_{L,T}(\omega,k)$ is the hadronic part of the electro-magnetic
vector current correlation function.  If the system evolves in a RHIC, it should
be folded with the evolution of the system.   

Assuming no change in the hadron properties at finite temperature or density,
the expected dilepton spectrum from a hot dense system will have a peak near 
the vector mesons.  Now, if  the vector mesons change their 
properties in 
medium, changes will occur in the peak.  The three characteristic changes we 
will discuss here are,  i) peak (mass) shift, ii) increased decay width, 
iii) dispersion effect (or finite ${\bf q}$) effect.

\subsection{Model calculation of Im$\Pi^V$ at finite temperature and 
density}

Many model calculations have been performed to calculate the changes of the 
vector meson properties at finite temperature or density.  
The models are either based on effective quark lagrangian or the 
effective hadron lagrangian.   The former approach have model dependence 
related to modeling the non-perturbative nature of QCD into the quark 
language.  The latter have large model dependence in extrapolating any 
effective hadron lagrangian to the on shell point of the vector meson mass 
region.  Table 3  summarizes the results of the model calculations.

\begin{table}[hbt]
\catcode`?=\active \def?{\kern\digitwidth}
\caption{ Model calculation for Im$\Pi_V$ at finite temperature and 
density.  The downward, upward and horizontal  arrows show  respectively 
the  decreasing , increasing and little change of values.  
 \label{tab:model}}
\begin{tabular*}{\textwidth}{@{}l@{\extracolsep{\fill}}rrrrr}
\hline
  & model  & mass  & width & dispersion \\ 
\hline
  & Walecka model \cite{wald} &  $\searrow$ & $\rightarrow$ & $\rightarrow$ \\
\cline{2-5}
 & Quark-meson coupling model\cite{qmd} & $\searrow$ & $\rightarrow$ & $\rightarrow$ \\
\cline{2-5}
Finite & Phenomenological meson exchange\cite{friman1} & $\searrow$ & $\rightarrow$ & $\rightarrow$ \\
\cline{2-5}
density & Brown-Rho scaling \cite{br} & $\searrow$ & $\rightarrow$ & $\rightarrow$  \\
\cline{2-5}
 & $\pi-N-\Delta$ dynamics\cite{pndd} &$\rightarrow$ & $\nearrow$  & $\rightarrow$ \\
\cline{2-5}
 & $V-N-N^\ast$ dynamics\cite{vnnd} & $\rightarrow$  & $\nearrow$ & $\nearrow$ \\
\hline
   & Chiral model with Vector mesons\cite{KG91,song92} &  $\nearrow$ & $\rightarrow$ & $\rightarrow$ \\
\cline{2-5}
 & Chiral model with Low T theorem\cite{LSY} & $\rightarrow$ & $\rightarrow$&  $\rightarrow$ \\
\cline{2-5}
Finite & $\sigma$-model with VMD\cite{Pisarski95} & $\nearrow$ & $\rightarrow$& $\rightarrow$ \\
\cline{2-5}
temperature & Hidden local symmetry\cite{Harada97} & $\nearrow$ &$\rightarrow$ & $\rightarrow$ \\
\cline{2-5}
 & Collision broadening\cite{Haglin95} & $\rightarrow$ & $\nearrow$ &$\rightarrow$ \\
\cline{2-5}
 & Massive YM with baryons\cite{Ko95} & $\searrow$ & $\rightarrow$ & $\rightarrow$ \\
\hline
\end{tabular*}
\end{table}

\subsection{QCD constraints on ${\rm Im}\Pi^V$}

Let us now try to construct QCD constraints on ${\rm Im}\Pi^V$ emphasizing
its connection to chiral symmetry restoration.    Consider either 
the transverse 
or the longitudinal polarization function in Eq.(\ref{pol2}).  
The starting point is the 
energy dispersion relation at finite ${\bf q}$.  For small ${\bf q}^2<\omega^2
$, we 
can make a Taylor expansion of the correlation function, which is even function
of $\omega$ and ${\bf q}$, such that,
 \begin{eqnarray}
\label{ope1}
{\rm Re} \Pi_{L,T}(\omega,{\bf q}) &  = & {\rm Re} \left( 
\Pi^0(\omega,0)+ 
\Pi_{L,T}^1(\omega,0) ~ {\bf q}^2 + \cdot \cdot \right) \nonumber \\  
 & = & \frac{1}{\pi} \int_0^\infty du^2 \left(
 { {\rm Im}\Pi^0(u,0) \over (u^2-\omega^2)} +  { {\rm Im}\Pi_{L,T}^1(u,0) \over (u^2-\omega^2)} ~ {\bf q}^2 + \cdot \cdot \right) , 
\label{dis1}
\end{eqnarray}
where we have assumed the retarded correlation function.  

The real part of eq.(\ref{dis1})  is calculated via 
the Operator Product Expansion (OPE) at large $-\omega^2 \rightarrow 
\infty$ with  finite ${\bf q}$.  The full polarization tensor will have the following form.
\begin{eqnarray}
\label{ope2}
\Pi_{\mu \nu}(\omega,{\bf q})  &  =  & 
(q_\mu q_\nu-g_{\mu \nu}) 
\left[- c_0 {\rm ln}|Q^2|+ \sum_d {c_{d,d} \over Q^d} A^{d,d}(n.m.) \right] 
\nonumber \\[12pt]
 & & +  \sum_{s,\tau=2} {1 \over Q^{s+\tau-2}} 
\left[ c_{d,\tau} 
g_{\mu \nu}  q^{\mu_1} \cdot \cdot q^{\mu_s} A^{d,\tau}_{\mu_1 ..\mu_s}
(n.m.) + \cdot \cdot  \right],
\end{eqnarray}
where, $Q^2={\bf q}^2-\omega^2$. 
Here, $A^{d,\tau}(n.m.)$ represents the nuclear matter expectation 
value of an operator of 
 dimension $d$ and twist $\tau=d-s$, where $s$ is the number of spin 
index.  These operators are defined at the 
scale  $Q^2$ and the $c$'s are the 
dimensionless Wilson coefficients  with the running coupling constant.   
This way of including the density 
effect is consistent at low energy\cite{HKL93}. 
The first set of terms in eq.(\ref{ope2}) 
come from the OPE of 
scalar operators, the second set from  operators with non zero spin $s$.
From this general expression, we can extract the OPE of $\Pi^0$, 
 $\Pi^1_{L,T}$.    Including  the contribution up to  dim.6 operators
only, they will have the following form,
\begin{eqnarray}
\label{ope3}
\Pi^0(\omega) &= &\Pi_{pert}(\omega)+\frac{1}{\omega^4} b_4^0+ \frac{1}{\omega^6} b_6^0
 \nonumber \\
\Pi^1(\omega) & = & \frac{1}{\omega^6} b_6^1+ \frac{1}{\omega^8} b_8^1
\end{eqnarray}
where $b_4^0 (b_6^0)$ in $\Pi^0$ are contributions from dim.4 (dim.6) operators and
 $b_6^1 (b_8^1)$ in $\Pi^1$ are from dim.4 (dim.6) operators.  
 It should be noted that in 
calculating $\Pi^1$,  
 it is enough to calculate only the ``non-trivial" ${\bf q^2}$ dependence
; i.e. neglect the dependence coming from expanding $Q^2={\bf q^2}-\omega^2$. 
That is so because we are interested only in the Lorentz breaking effect and not
 in the trivial ${\bf q^2}$ dependence\cite{Lee97}.
 Eq.(\ref{ope3}) 
are asymptotic expansions in $1/\omega^2$.  Looking back at the 
dispersion relation in Eq.(9), we can look at different moments of the 
imaginary part and equate it to each term in the OPE in Eq.(\ref{ope3}).  

\begin{eqnarray}
\label{constraint1}
\int du^2  \left( {\rm Im}\Pi(u)-{\rm Im}\Pi_{pert}(u) \right) u^{n-2}  =  -b_n  
\end{eqnarray}
Here, ${\rm Im}\Pi(u)_{pert}$ is the corresponding imaginary part of $\Pi_{pert}$.  
 For ${\rm Im}\Pi^0$, we have 3 constraints from $n=2,4,6$ with $b_2=0$, and 
for 
 ${\rm Im}\Pi^1$, we have 3 constraints from $n=4,6,8$ with $b_4=0,{\rm Im}\Pi_{pert}=0$  
for both the 
longitudinal and transverse directions.   These constraints are model 
independent constraints and relates QCD perturbative calculations and 
expectation values of local operators to the phenomenologically 
observable spectral density ${\rm Im}\Pi(u)$.   The usefulness of the constraints
can be summarized as follows.

\begin{enumerate}

\item   The  temperature dependence (density dependence) of the local 
 condensates $(b's)$ can be calculated using lattice QCD.  Therefore, 
starting from the vacuum form of ${\rm Im}\Pi(u)$, one can study its changes 
at finite temperature or density.  

\item One can look at the effect of each local operators in $b's$ separately.  
 Hence, from looking at  the effect of chiral symmetry breaking operators, 
 one can study the effect of chiral symmetry restoration on Im$\Pi(u)$.  

\item All model calculations of ${\rm Im}\Pi(u)$, or even the experimental data
itself should satisfy the above constraints.   

\end{enumerate}

It is straightforward to extract the chiral symmetry breaking operators.  
 First, it should be noted that all the twist-2 and Gluonic operators, appearing in Eq.(11), are 
 chirally symmetric operators.   Four quark operators have a chirally 
symmetric component and  a breaking part.  Assuming a  four quark operator 
 ${\cal O}$,  
one can divide these parts by subtracting and adding its chiral partner
${\cal O}=\frac{1}{2} [ {\cal O}  -{\cal O}({\rm chiral~ partner}) ]
+\frac{1}{2} [ {\cal O}  + {\cal O}({\rm chiral~ partner}) ] $.  
 The explicit form of $b's$ can be found in \cite{HKL93} for 
 Im$\Pi^0$ and in \cite{Lee97} for Im$\Pi^1$.  
It would be extremely useful to study the temperature dependence of these 
chiral symmetry breaking operators on the lattice and study its 
phenomenological consequences using these constraints.  

\section{QCD SUM RULE RESULT FOR ${\rm Im}\Pi^V$}

We will now discuss specific changes of ${\rm Im}\Pi^V$ in the QCD sum rule
approach.  

\subsection{ Borel sum rule for $\Pi^0$ at finite density}

After the Borel transformation with respect to $\omega$, the  dispersion relation in Eq.(9) with the OPE in 
Eq.(\ref{ope3}) looks as follows.

\begin{eqnarray}
\int[ {\rm Im}\Pi^0(u)-{\rm Im}\Pi^0(u)_{pert}(u) ]e^{-u^2/M^2} du^2=
\sum_{n=2,4} \frac{(-1)^{n/2+1}}{(n/2)! M^n}b_n^0
\end{eqnarray}

Motivated by its form in the vacuum, 
the spectral density will be modeled with a pole and a continuum.
\begin{eqnarray}
{\rm Im}\Pi^0(u)= { F \Gamma  \over (u^2-m_V^2)^2+ (m_V \Gamma )^2 }
+ c \theta(u^2-s_0) + \rho_{scattering}
\end{eqnarray}
The Borel transformation will reduce the uncertainty associated with 
approximating the continuum with a sharp step function.  Nevertheless, it 
gives a maximum Borel Mass $M^2_{max}$ above which the continuum contribution
becomes too large.   $\rho_{scattering}$ is the possible subtraction constants
proportional to $\delta(u^2)$, which can be calculated using the nucleon-hole 
contribution\cite{HL92,HSL95}.   

The next step is to calculate the changes in the condensate.  We will use the 
linear density approximation.  In this limit, the nuclear matter expectation 
values of any local operator ${\cal O}$ can be approximated as ,
\begin{eqnarray}
\langle {\cal O} \rangle_{n.m.}= \langle {\cal O} \rangle_0
+ n_n \langle {\cal O} \rangle_N,
\end{eqnarray}
where $ \langle \cdot \rangle_0$ is the vacuum expectation value, 
 $\langle \cdot \rangle_N$ is the nucleon expectation value and $n_n$ is the 
baryon density.  
Up to dimension 6, all the relevant expectation values appearing in the $\Pi^0$ 
sum rules are known except for the dim. 6 four quark operator, for which we
 will use the mean field approximation\cite{HL92,HSL95}.  
The truncation at dim 6 operator determines the minimum Borel mass 
 $M^2_{max}$, which is determined by assuming that the power corrections are 
less than 30\% of the perturbative contribution.  

The final step is to allow the phenomenological parameters to change also 
in density and determine its change by making a best fit of the 
left hand side to the right hand side of  Eq.(13)  between the Borel interval $M^2_{min}$ and 
$M^2_{max}$.  

This was first   
applied to the vector mesons  by Hatsuda and Lee \cite{HL92} 
 in the limit $\Gamma \rightarrow 0$.  The vector meson peak was shown to shift 
as follows,
\begin{eqnarray}
\frac{ m_V(n_n)}{m_V(0)} & = & 1- (0.16 \pm 0.06) \frac{n_n}{n_0}, \nonumber 
\\
\frac{ m_\phi(n_n)}{m_\phi(0)}& = & 1- (0.03 \pm 0.015) \frac{n_n}{n_0},
\end{eqnarray}
where the first line denotes the result for the $\rho$ and $\omega$
 meson and the 
second line that for the $\phi$ meson. $n_0$ denotes the nuclear saturation 
density.  The result for the $\rho,\omega$ 
meson were due mainly to the changes in the four quark condensate
 $\langle ({\bar q} q)^2 \rangle$ and about 30\% 
from  $\langle {\bar q} \gamma_\mu D_\nu q 
\rangle$.  The former in the mean field approximation is a chiral symmetry 
breaking operator and the latter is a chirally symmetric operator.  
As for the $\phi$ meson, the effect comes mainly from the chiral symmetry 
breaking operator $\langle m_s {\bar s} s \rangle$.    Therefore, the shift in 
masses come mainly from chiral symmetry restoration.

Recently, the work has been extended to include possible change in the 
width\cite{Mosel97}.  It was found that the downward shift of the mass 
could be compensated by an increase in the width.  Hence, there is a band
of region in the mass vs. width plane where the sum rule in Eq.(13) is reasonably 
satisfied.   However, there exist a best fit, which starting from the 
vacuum value changes as follows for the $\rho,\omega$ meson,
\begin{eqnarray}
\label{vector}
\Delta m_V \sim -50  ~\frac{n_n}{n_0}~ {\rm MeV}~,~~
\Delta \Gamma \sim + 200  ~\frac{n_n}{n_0}~ {\rm MeV}.
\end{eqnarray}
Similar analysis are in progress for the $\phi$ meson.  

Summarizing the results on $\Pi^0$, QCD sum rule suggests that due to 
chiral symmetry restoration, there will be a decrease  in mass and 
increase in the width.

\subsection{ Borel sum rule for $\Pi^1$ at finite density}

As shown in Eq.(8), the dilepton spectrum will have contribution from 
both the longitudinal and transverse polarization function, which will 
now be  a function of both $\omega$ and ${\bf q}$.  We 
will allow the parameters in Eq.(14) to vary non-trivially by a term 
proportional to ${\bf q^2}$ and implicitly the nuclear density $n_n$.    
Such as, 
\begin{eqnarray}
F \rightarrow F+f \cdot {\bf q^2},~~m_V^2 + a \cdot {\bf q^2},~~
S_0 \rightarrow S_0 +s \cdot {\bf q^2}.
\end{eqnarray}
In the limit $\Gamma \rightarrow 0$, this will give the following form for
 Im$\Pi^1$,
\begin{eqnarray}
{\rm Im}\Pi^1(u) = f \cdot \delta(u^2-m_V^2)-a \cdot F \delta'(u^2-m_V^2)
 -s \cdot c_0 \delta(u^2-S_0)+8 \pi^2 n_n b_{scatt} \delta'(u^2).
\label{phen2}
\end{eqnarray} 
Here, we have included the subtraction constant associated with the 
 $\delta'(u^2)$ singularity, which can be calculated from the 
nucleon-hole contribution.   There is another unknown
 $\delta(u^2)$ singularity.  To eliminate this, 
 we use a once subtracted dispersion relation for $\Pi^1$.
\begin{eqnarray}
\int[ {\rm Im}\Pi^1(u)-{\rm Im}\Pi^1(u)_{pert}(u) ]u^2 e^{-u^2/M^2} du^2=  \sum_{n=2,4} \frac{(-1)^{n/2}}{(n/2-1)! M^{n+2}}b_n^1
\end{eqnarray}
 From a best fit analysis, one can extract the momentum dependent parameters
\cite{Lee97}.  The results are shown in Table  4.  

\begin{table}[hbt]
\catcode`?=\active \def?{\kern\digitwidth}
\caption{   Results for the parameters at nuclear matter density.  The
values are from best fit of the Borel sum rule in eq.(20).
\label{tab:qeffect}}
\begin{tabular*}{\textwidth}{@{}l@{\extracolsep{\fill}}rrrr}
 \hline 
 & $a$ & $f$ & $s$  \\ \hline 
Transverse $\rho$ & -0.108 & 0.190 & -0.028 \\
Transverse $\omega$ & -0.081 & 0.171 & -0.010  \\
Longitudinal $\rho,\omega$ & 0.023 & 0.066 & 0.029 
 \\
Transverse $\phi$ & 0.004 & 0.010 & 0.009 \\
Longitudinal $\phi$ & 0.009 & -0.001 &
0.009  \\
\hline
\end{tabular*}
\end{table}

 As discussed before, a non-vanishing $a$ will shift the average 
 peak position by 
 $\Delta M= \sqrt{m_V^2+a {\bf q}^2}-m_V$, even if there is no change in the
scalar mass $m_V$.  With the values of $a$ obtained, we find that 
 even at  ${\bf q} = .5$GeV/c, 
above which point our formalism breaks down, the shifts are less than 2\%  (0.05 \%) at 
nuclear matter density for the $\rho,\omega$ ($\phi$).  This is  smaller than 
the expected 
scalar mass shift of the $\rho,\omega$ ($\sim$ 20\%) and 
 $\phi$ ($\sim$3\%) \cite{HL92} and 
justifies (to a first approximation) neglecting the three momentum dependence and the polarization effect
 when implementing the universal scaling laws (Brown-Rho scaling)
  of the vector mesons in understanding 
 the dilepton spectrum in A-A and p-A reaction\cite{LKB95}.

The contribution of the longitudinal and transverse polarization to the
 dilepton 
spectrum, depends on the angle between the sum and difference of the 
three momentum of the out going dileptons\cite{KG91}.  
However, after averaging, 
the contribution of the transverse polarization becomes twice that of the 
longitudinal polarization, as can be seen from Eq.(8). 
 Hence, to a  good approximation, one can implement the finite ${\bf q}$ effect 
 into model calculations by including only the transverse 
dispersion relation.  Making a linear fit at the nuclear matter density, one can parameterize the 
vector meson mass in medium as follows,
\begin{eqnarray}
\frac{m_\rho(n_n)}{m_\rho(0)} & = & 1-(0.023 \pm 0.007) \left( \frac{ {\bf q}}{0.5} \right)^2 \frac{n_n}{n_0} 
\nonumber \\
\frac{m_\omega(n_n)}{m_\omega(0)} & = & 1-(0.016 \pm 0.005) \left( \frac{ {\bf q}}{0.5} \right)^2 \frac{n_n}{n_0}
\nonumber \\
\frac{m_\phi(n_n)}{m_\phi(0)} & = & 1 
+(0.0005 \pm 0.0002) \left( \frac{ {\bf q}}{0.5} \right)^2 \frac{n_n}{n_0},
\end{eqnarray}
where ${\bf q}$ is in the GeV unit.  In a full model calculation, the 
effect in Eq.(21) should be added to that of Eq.(16).  
It should be noted that the main operators responsible for the result in 
Table 4 come mainly from chirally symmetric twist-2 operators.  So the 
dispersion effect is a chirally symmetric effect and has little 
relation to chiral symmetry restoration.  

\section{SUMMARY}

\begin{enumerate}

\item Chiral symmetry is restored at high temperature and/or density, while
 $U_A(1)$ breaking effect persist above the transition.  It would be useful 
to perform lattice calculations of $U_A(1)$ breaking effect above the chiral 
phase transition using 2 quark-bilinear order parameter for quark
flavors larger than 2.

\item Using the QCD constraints  one can relate observed changes in spectral density to 
changes of local operators at finite temperature or density.  A systematic
lattice gauge theory calculations on the temperature dependence of the chiral symmetry breaking condensate would be a useful future project.  

\item QCD sum rules in matter suggests that the vector meson 
width will increase and its peak position shift downwards in energy at finite 
density and temperature before it becomes part of the continuum above 
the critical boundary.  These changes are due mainly to chiral 
symmetry restoration. 

\item  QCD sum rule in matter suggests that the dispersion effect (
finite ${\bf q}$ effect) is small, but should be  included in a full
study of the changes in the vector meson peak in medium.  

\end{enumerate}


\section{ACKNOWLEDGEMENTS}

I would like to thank the organizers for the invitation to give a talk.    
I would like to thank 
T. Hatsuda for a close collaboration on the presented subject and  
T. D. Cohen, B. Friman and Hungchong Kim for various useful discussions.


\begin{thebibliography}{99}

\bibitem{Oka} D. Jido, N. Kodama, M. Oka, \Journal{\PRD}{54}{4532}{1996}.

\bibitem{Hung} Su H. Lee and H. Kim , \Journal{\NPA}{612}{418}{1997}.

\bibitem{Columbia} F.R. Brown {\it et al}, \Journal{\PRL}{65}{2491}{1990}.

\bibitem{Iwasaki} Y. Iwasaki {\it et al}, \Journal{\ZPC}{71}{343}{1996}.

\bibitem{Schafer} T. Sch\"{a}fer, hep-ph/9708256.

\bibitem{random} T. Wettig, T. Guhr, A. Schafer, H. A. Weidenmuller, hep-ph/9701387.

\bibitem{Bernard97} C. Bernard {\it et al},  \Journal{\PRL}{78}{598}{1997}.

\bibitem{Boyd} G. Boyd, hep-lat/9607046.

\bibitem{tHooft} G. 't Hooft, , \Journal{\PRL}{37}{8}{1976}:
\Journal{\PRD}{14}{3432}{1976}, M.A. Shifman, A.I. Vainshtein, V.I. Zakharov, 
\Journal{\NPB}{163}{46}{1980}.

\bibitem{LH96} Su H. Lee and T. Hatsuda,  \Journal{\PRD}{54}{r1871}{1996}

\bibitem{BS86} A.I. Bochkarev and M.E. Shaposhnikov, \Journal{\NPB}{268}{220}{1986}

\bibitem{Pi82} R. Pisarski, \Journal{\PLB}{110}{222}{1982}


\bibitem{wald} H.-C. Jean, J. Piekarewicz, A.G. Williams, 
\Journal{\PRC}{49}{1981}{1994};H. Shiomi and T. Hatsuda, \Journal{\PLB}{334}{281}{1994}.

\bibitem{qmd} K. Saito and A. W. Thomas, \Journal{\PRC}{51}{2757}{1995}.

\bibitem{friman1} B. Friman, M. Soyeur, \Journal{\NPA}{600}{477}{1996}.

\bibitem{br} G.E. Brown and M. Rho, \Journal{\PRL}{66}{2720}{1991}.

\bibitem{pndd} M. Asakawa, C.M.Ko, P. Levai and X.J. Qiu, \Journal{PRC}{46}{1157}{1992};
G. Chanfray and P. Schuck, \Journal{NPA}{545}{271}{1992};
M. Herrmann, B.L.Friman and W. N\"{o}renberg, \Journal{\NPA}{560}{411}{1993};
F. Klingl, N. Kaiser and W. Weise, \Journal{\NPA}{624}{527}{1997}.


\bibitem{vnnd} B. Friman, H.J. Pirner, \Journal{\NPA}{617}{496}{1997};
R.Rapp, G. Chanfray and J. Wambach, \Journal{\NPA}{617}{472}{1997};
W. Peters {\it et al} nucl-th/9708004.


\bibitem{KG91} C. Gale and J. I. Kapusta, \Journal{\NPB}{357}{65}{1991}.


\bibitem{song92}  C. Song, \Journal{\PRD}{48}{1375}{1993}.

\bibitem{LSY} Su H. Lee, C. Song and H. Yabu, \Journal{\PLB}{341}{407}{1995}.

\bibitem{Pisarski95}  R. D. Pisarski, \Journal{\PRD}{52}{3773}{1995}.

\bibitem{Harada97} M. Harada and A. Shibata, \Journal{\PRD}{55}{6716}{1997}.

\bibitem{Haglin95}  K. L. Haglin, \Journal{\PRC}{54}{1492}{1996}; \Journal{\NPA}{584}{719}{1995}; nucl-th/9710026.

\bibitem{Ko95} C. Song, P.W. Xia and C.M. Ko, \Journal{\PRC}{52}{408}{1995}

\bibitem{HKL93} T. Hatsuda, Y. Koike and Su H. Lee, \Journal{\NPB}{394}{221}{1993}.

\bibitem{Lee97} Su H. Lee, nucl-th/9705048.

\bibitem{HL92} T. Hatsuda and Su H. Lee, \Journal{\PRC}{46}{r34}{1992}.

\bibitem{HSL95} T. Hatsuda, Shiomi and Su H. Lee, \Journal{\PRC}{52}{3364}{1995}.

\bibitem{Mosel97}  S. Leupold, W. Peters and U. Mosel, nucl-th/9708016.

\bibitem{LKB95} G.Q.Li, C.M. Ko and G.E.Brown,  \Journal{\PRL}{72}{4007}{1995}.

\end{thebibliography}
\end{document}